\documentclass[traditabstract,letter]{aa} 
\usepackage{graphicx}
\usepackage{natbib}
\usepackage{txfonts}

\newcommand{\prep}[1]{#1}
\newcommand{\bfx}[1]{#1}

\renewcommand{\deg}{^{\circ}}
\newcommand{\sta}{{\star}}

\newcommand{\esc}{{\rm es}}
\newcommand{\Jup}{{\rm Jup}}
\newcommand{\Nep}{{\rm Nep}}
\newcommand{\Sat}{{\rm Sat}}
\newcommand{\PDF}{\mathop{\rm PDF}\nolimits}

\newcommand{\vect}[1]{{\hat{\vec{#1}}}}

\def\norm#1{\left\Vert#1\right\Vert}
\newcommand{\moy}[2]{\left\langle{#2}\right\rangle_{#1}}

\def\crm{\cr\noalign{\medskip}}

\def\m@th{\mathsurround=0pt}
\def\EQM#1{\vcenter{\normalbaselines\m@th
    \ialign{${\displaystyle ##}$\hfil&&\ ${\displaystyle ##}$\hfil\crcr
    \mathstrut\crcr\noalign{\kern-\baselineskip}
    \noalign{\smallskip}
    #1\crcr\mathstrut\crcr\noalign{\kern-\baselineskip}}}}

\newcommand{\be}{\begin{equation}}
\newcommand{\ee}{\end{equation}}

\newcommand{\bpm}{\left(\begin{array}{c}}
\newcommand{\epm}{\end{array}\right)}

\newcommand\captiona{
 Geometry of evaporation.
}

\newcommand\figa{
\begin{figure}
\begin{center}
\includegraphics[width=5cm]{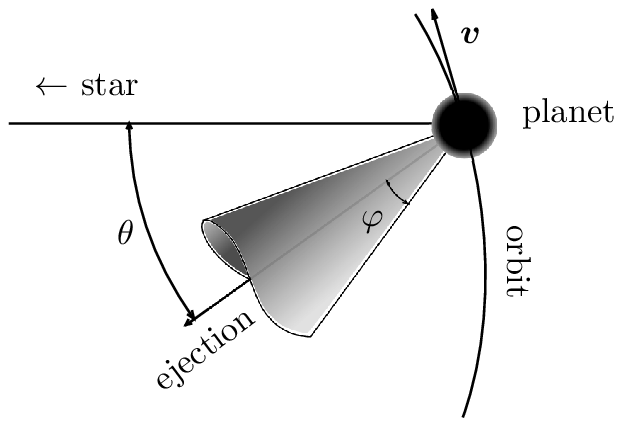}
\caption{\captiona}
\label{Figa}
\end{center}
\end{figure}
}

\newcommand\captionb{(a)
Distribution of orbital periods for all planets with $m < 5$
$M_\Jup$ in white, Jupiters in gray, and Neptunes in black. (b)
CDFs with their best fit by a log-normal distribution. Jupiters
are represented by triangles, and Neptunes by circles.}

\newcommand\figb{
\begin{figure}
\begin{center}
\includegraphics[width=7cm]{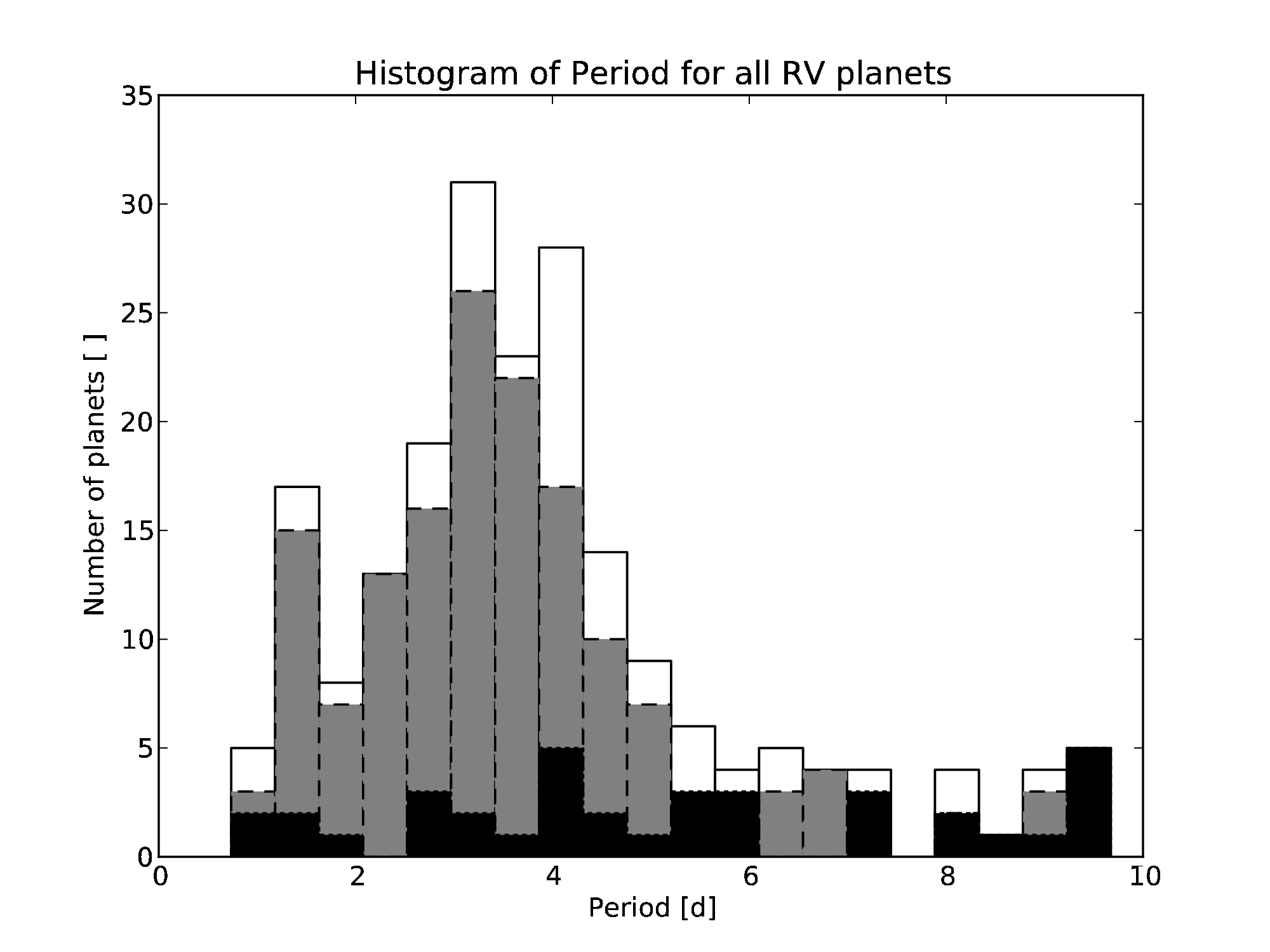}
\includegraphics[width=7cm]{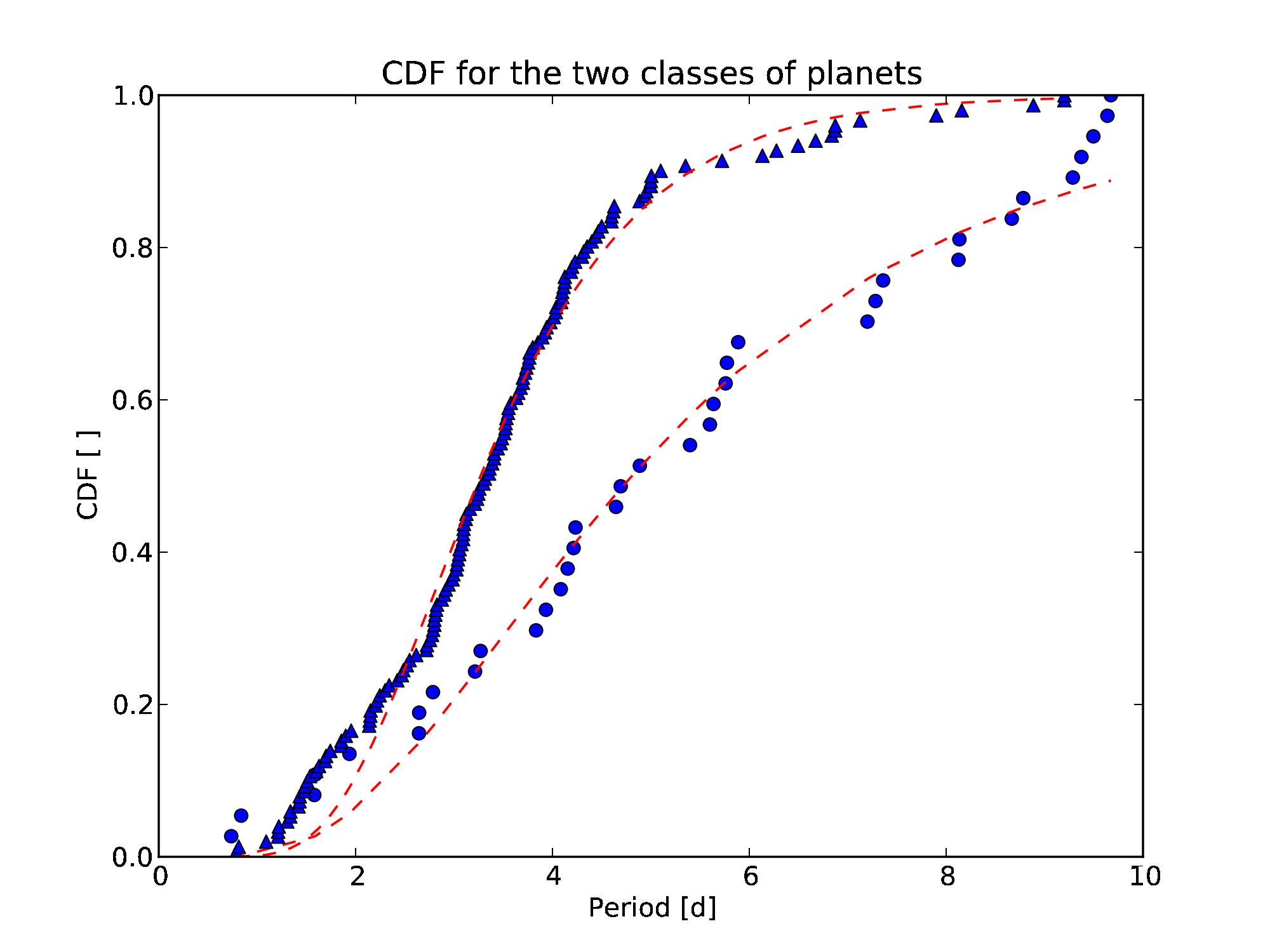}
\caption{\captionb}
\label{fig_histo}
\end{center}
\end{figure}
}

\newcommand\captionc{PDF of the migration efficiency $\tau$ as a function of
$\tau$ (a), or as a function of the angles $\theta$ and $\varphi$
(b). In (b), the solid line corresponds to the maximum likelyhood, the dashed lines
delineates the one sigma confidence region, and the dotted line, the two
sigma region. The small circle, at longitude 30$\deg$ and
aperture 30$\deg$, shows the approximate location of the hottest
region of HD 189733b \citep{Knutson_etal_nature_2007}.}

\newcommand\figc{
\begin{figure}
\begin{center}
\includegraphics[width=7cm]{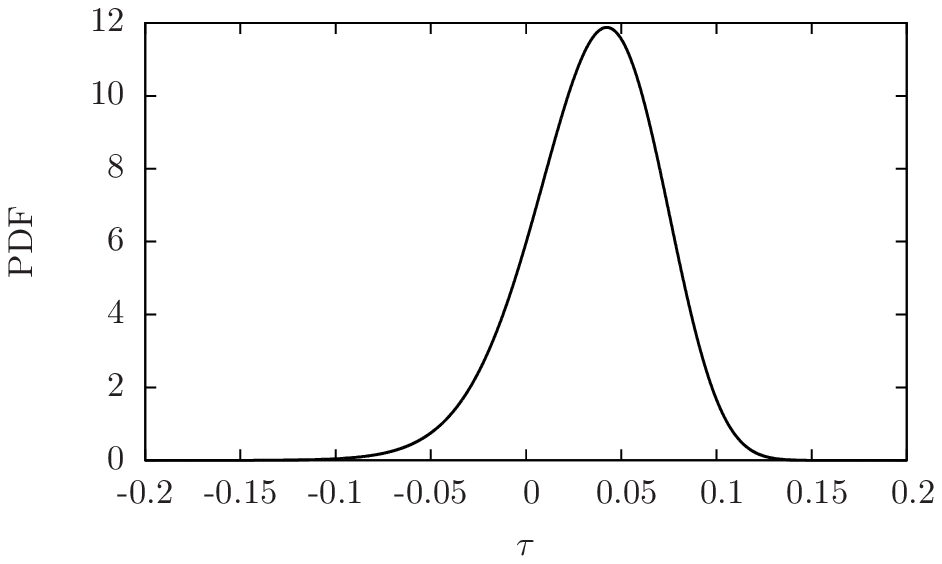}
\includegraphics[width=7cm]{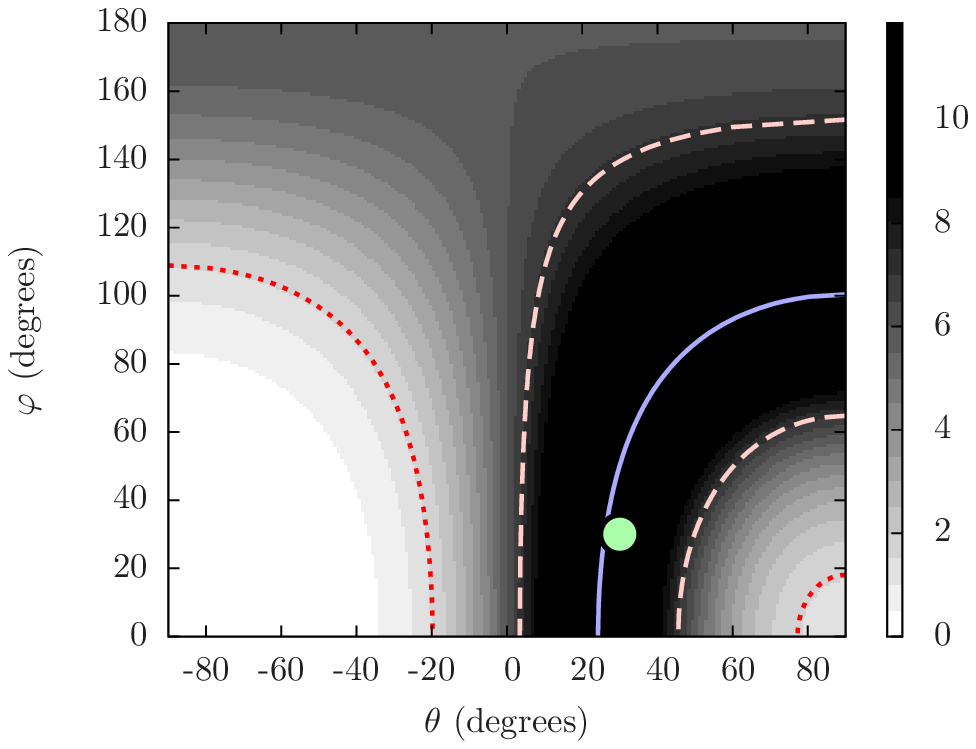}
\caption{\captionc}
\label{fig_pdftau}
\end{center}
\end{figure}
}

\begin{document}
   \title{Orbital migration induced by anisotropic evaporation}

    \subtitle{Can hot Jupiters form hot Neptunes ?}

   \author{G. Bou\'e
          \inst{1,2}
          \and
          P. Figueira
          \inst{1}
          \and
          A.C.M. Correia
          \inst{2,3}
          \and
          N.C. Santos
          \inst{1,4}
          }

   \institute{Centro de Astrof\'isica da Universidade do Porto, Rua das
              Estrelas, 4150-762 Porto, Portugal \\
              \email{gwenael.boue@astro.up.pt}
         \and
              ASD, IMCCE-CNRS UMR8028, Observatoire de Paris, UPMC, 77
              avenue Denfert-Rochereau, 75014 Paris, France
         \and
              Department of Physics, I3N, University of Aveiro, Campus
              Universit\'ario de Santiago, 3810-193 Aveiro, Portugal
         \and
              Departamento de F\'isica e Astronomia, Faculdade de
              Ci\^encias, Universidade do Porto, Portugal              
             }

   \date{Received ..., ...; accepted ..., ...}

  \abstract
{Short period planets are subject to intense energetic irradiations from
their stars. It has been shown that this can lead to significant
atmospheric mass-loss and create smaller mass planets. Here, we analyse
whether \bfx{the evaporation mechanism can affect the orbit of planets.}
The orbital evolution of a planet undergoing evaporation is derived
analytically in a very general way. Analytical results are then compared
with the period distribution of two classes of inner exoplanets:
Jupiter-mass planets and Neptune-mass planets. These two populations
have a very distinct period distribution, with a probability lower than
$10^{-4}$ that they were derived from the same parent
distribution. We show that mass ejection can generate significant
migration with an increase of orbital period that matches very well the
difference of distribution of the two populations. This would happen if
the evaporation emanates from above the hottest region of planet
surface.  Thus, migration induced by evaporation is an important
mechanism that cannot be neglected.}


   \maketitle
%

\section{Introduction}
Different scenarios have been proposed to explain the formation of hot
Neptunes and Super-Earths: migration in a protoplanetary disk
\citep{Mordasini_etal_AA_2009}, in situ formation by accretion of
planetesimals \citep{Brunini_Cionco_Icarus_2005}, embryo formation in a
compact multiplanetary system and subsequent migration through
scattering \citep{Ida_Lin_ApJ_2010}, tidal downsizing of giant planet
embryo \citep{Nayakshin_MNRAS_2010, Nayakshin_MNRAS_2011}, or partial
evaporation of more massive planets \citep{Baraffe_etal_AA_2004,
Lecavelier-des-Etangs_AA_2007, Valencia_etal_AA_2010}. Here, we focus on
the evaporation hypothesis with a particular attention to semi-major
axis evolution. Indeed, among the shortest period exoplanets, those
closer to their star tend to be more massive than the outer ones
\citep{Mazeh_etal_MNRAS2005, Southworth_etal_MNRAS_2007,
Davis_Wheatley_MNRAS_2009MNRAS, Llambay_etal_AA_2011}. This has been
attributed either to a combined effect of tidal interactions and
migration in the protoplanetary disk \citep{Llambay_etal_AA_2011}, or to
the fact that low-mass close-in planets cannot survive catastrophic
evaporation \citep{Davis_Wheatley_MNRAS_2009MNRAS}. 

While isotropic planetary evaporation generates almost neglectable
migration, anisotropic ejection of matter can lead to substantial
increase of semi-major axis. This effect is similar to anisotropic
thermal radiation \citep{Fabrycky_ApJL_2008}.  In the case of hot
gazeous planets, two different mechanisms have been proposed to
explain atmospheric escapes, i.e. how particules reach the Roche lobe
(Hill) radius of their planet. On the one hand, this may be due to a
thermal inflation of the planet. Particules of the upper atmosphere are
assumed to follow a Maxwellian velocity distribution, and those with the
highest speed escape \citep{Gu_etal_ApJ_2003,
Lecavelier-des-Etangs_etal_AA_2004}. In that case, the evaporation is
preferentially directed towards the L1 Lagrangian point located between
the star and the planet \citep{Gu_etal_ApJ_2003}. On the other hand, the
escape may be driven by vertical winds on the planet's dayside
\citep{Yelle_Icarus_2004, Garicia_PSS_2007, Murray-Clay_etal_ApJ_2009}.
This situation is more complex and 3D hydrodynamical simulations are
required to get the geometry of ejection. Nevertheless, winds are
likely anisotropic.

Atmospheric circulation models predict the existence of a jet that
displaces the hottest region of hot Jupiters from the substellar point,
along the equator \citep{Showman_Guillot_AA_2002}.  More recent models
agree that the jet is a common feature of circulation models, and that
it is confined to latitudes between -30$\deg$ and 30$\deg$ with a
displacement from the substellar point usually eastward, by 10-60$\deg$
of longitude. Its value depends on the strength of the imposed stellar
heating and other factors such as the radiative timescale and drag time
constant \citep{Showman_Polvani_arxiv_2011}. These predictions are
corroborated by the observed surface temperature distribution of HD
189733b \citep{Knutson_etal_nature_2007}. 

Although there are evidences of a displacement of the hottest
region with respect to the substellar point on planet surface, the
dynamics of exobase is poorly known. Strictly speaking, it is
thus not possible to assert in which direction and how mass is ejected
by evaporation. Hereafter, in order to get estimates of the
migration induced by evaporation, we will take the results mentionned
above concerning planetary surface as representative of the exobase 
structure.

\bfx{This letter is organized as follows. In Section~\ref{sec.orbit}, 
we derive analytically the orbital evolution of a planet
undergoing evaporation in a very general way. In
Section~\ref{sec.observation}, we compare the distribution of two
classes of close-in planets: Jupiter-like planets
and Neptune-like planets. The statistical analysis is then used, in
Section~\ref{sec.comparison}, to constrain the geometry of evaporation 
assuming that hot Neptunes are partially evaporated hot Jupiters.
This geometry is also confronted with the predictions of atmospheric
models. The conclusion is presented in Section~\ref{sec.conclusion}.}

\section{Orbital evolution with evaporation}
\label{sec.orbit}
Two-body problems in which one of the component loses mass should be
solved with caution. Several misanderstandings have already been found
in the litterature \citep{Plastino_Muzzio_CeMe_1992}, since the
equations of motion are not the same whether the ejection is isotropic
or not.  Here, we consider a simple model of evaporation that is well
suited for both isotropic and anisotropic mass-loss. 

Let a planet of mass $m(t)$ orbiting a star with mass $m_\sta$. The
orbit is characterized by a radius vector $\vec r$, an orbital velocity
$\vec v = \dot{\vec r}$, a semi-major axis $a$, an eccentricity $e$, an
orbital angular momentum $\vec \ell = \vec r \times \vec v$, and a
period $P=2\pi/n$, where $n$ is the mean motion defined by $G(m+m_\sta)
= n^2a^3$. $G$ is the gravitational constant. Let $\vec v_p$ and $\vec
v_\sta$ be respectively the velocity of the planet and of the star in a
fixed reference frame. The orbital velocity is then $\vec v = \vec
v_p - \vec v_\sta$. For any vector $\vec w$, we note $w=\norm{\vec w}$
its norm, and $\vect w=\vec w/w$ its unit vector. The frame $(\vect r,
\vect \theta, \vect k)$, where $\vect k = \vect \ell$ and $\vect \theta
= \vect k \times \vect r$, is used to defined the geometry of the mass
ejection, which is given by two parameters: the angle $\theta$ between
the star and the direction of mass ejection, and the aperture $\varphi$
of the stream (see Fig.~\ref{Figa}). In particular, the evaporation is
completely focused in one direction when $\varphi=0\deg$, and becomes
isotropic when $\varphi=180\deg$.  We note $\vect u=-\cos \theta\, \vect
r - \sin \theta\, \vect \theta$ the average direction of the stream. 
\prep{\figa}
Let $V_\esc$ be the ejection speed of atmospheric particules with
respect to the planet barycenter. Note that $V_\esc$ can vary
between $10$ and $20$ km.s$^{-1}$ for a planet orbiting a Main-Sequence
star, and a T Tauri star, respectively
\citep{Murray-Clay_etal_ApJ_2009}. 
The mean velocity $\vec V$ of the stream is $\eta(\varphi) V_\esc \vect u$,
where
\be
\eta(\varphi) = \frac{\int_0^\varphi \cos \varphi' \sin \varphi' \,
d\varphi'}{\int_0^\varphi \sin \varphi'\, d\varphi'} = 
\cos^2\frac{\varphi}{2}\ ,
\ee
and the equations of motion are
\be
\EQM{
m\frac{d\vec v_p}{dt} &=& -G m_\sta m \frac{\vec r}{r^3} + \dot{m} \vec
V \ , \crm
m_\sta \frac{d\vec v_\sta}{dt} &=& +G m_\sta m \frac{\vec r}{r^3}\ .
}
\label{eq.motion}
\ee
We then assume that the mass-loss rate, $\dot m(t)$, scales linearly
with the energy received by the planet from the stellar XEUV emission.
It should be stressed that this has no effect on semi-major
axis. Another scaling would modify only slightly the eccentricity 
evolution. The mass
loss-rate is thus inversely proportional to the square of the distance
\be
\dot m(t) = - B(t)/r^2\ ,
\label{eq.mdot}
\ee
where $B(t)$ is any slowly varying function accounting for the time
evolution of the stellar luminosity. From (\ref{eq.motion}),
(\ref{eq.mdot}), and the expression of $\vec V$, one gets
\be
\frac{d \vec v}{dt} = -\mu(t)\frac{\vect r}{r^2} + \kappa(t) \sin \theta
\frac{\vect \theta}{r^2}\ ,
\label{eq.dvdt}
\ee
where $\kappa(t) = \eta(\varphi) B(t) V_\esc / m(t)$ and $\mu(t) =
G(m(t)+m_\sta)-\kappa(t) \cos \theta$. Notice that even in
the extreme case with a mass-loss rate $\dot{m} = -10^{15}$ g.s$^{-1}$,
and an ejection speed $V_\esc=100$ km.s$^{-1}$, for a hot Jupiter at
0.05 AU, $\kappa/(Gm)$ remains lower than $2\times 10^{-8}$. In the
following, we thus use simply $\mu(t) = G(m(t)+m_\sta)$.

From the definition of the semi-major axis
\be
\frac{1}{2}v^2 - \frac{\mu}{r} = -\frac{\mu}{2a}\ ,
\ee
and the expression of the
Laplace-Runge-Lenz vector $\vec e$
\be
\vec e = \frac{\vec v \times \vec \ell}{\mu} - \vect r\ ,
\ee
the equation of motion (\ref{eq.dvdt}) leads to
\be
\EQM{
\frac{da}{dt} &=& a\left(1-2\frac{a}{r}\right)\frac{\dot{\mu}}{\mu} +
2\frac{a^2 \ell \kappa \sin\theta}{\mu r^3}\ , \crm
\frac{d\vec e}{dt} &=& -\left(\vect r+\vec e\right)\frac{\dot{\mu}}{\mu}
+\frac{\kappa \sin\theta}{\mu r^3}(\ell \vec r+r^2 \vec v \times \vect
k)\ .
}
\label{eq.dade}
\ee
The first terms of the right-hand sides of Eq.~(\ref{eq.dade}) are only
due to the mass decrease. They do not depend on the geometry and remain
unchanged even for isotropic evaporation. On the other hand, the last
terms of Eq.~(\ref{eq.dade}) vanish for isotropic ($\kappa=0$) or radial
($\sin \theta=0$) mass ejection. The long term evolution of the orbital
parameters is obtained by averaging (\ref{eq.dade}) over the mean
anomaly $M$ keeping $\mu(t)$, $\kappa(t)$ and $\ell=na^2\sqrt{1-e^2}$
constant \citep[e.g.][]{Boue_Laskar_Icarus_2009}. The results are
\be
\EQM{
\moy{M}{\frac{da}{dt}} &=& -a\frac{\dot{\mu}}{\mu} +
2\frac{na\kappa\sin\theta}{\mu(1-e^2)}\ , \crm
\moy{M}{\frac{d\vec e}{dt}} &=&
\frac{n\kappa\sin\theta}{\mu(1+\sqrt{1-e^2})} \vec e\ .
}
\label{eq.secular}
\ee
In the following, we consider the simplest case where the orbit is
circular. Eq.~(\ref{eq.secular}) shows that such orbits remain circular
over the whole evolution. Replacing $B(t)$ (see Eq.~(\ref{eq.mdot})) by
$-a^2 \dot{m}(t)$ in the expression of $\kappa(t)$, and using $v=na$ and
$\mu(t) = G(m(t)+m_\sta)$, the secular evolution of the semi-major axis
becomes
\be
\moy{M,\,e=0}{\frac{da}{dt}} = -\frac{\dot m}{m_\star + m}a
-2 \tau \frac{v_0}{v}\frac{\dot m}{m}a\ ,
\label{eq_dadt}
\ee
where 
$v_0=v(t=0)$ is the initial orbital velocity. $\tau$ is a dimensionless
parameter that controls the efficiency of the migration. We will thus
refer to this parameter as the `migration efficiency'. Its expression
is
\be
\tau = \cos^2\frac{\varphi}{2}\sin\theta\frac{V_\esc}{v_0}\ .
\label{eq.tau}
\ee
For isotropic evaporation ($\varphi=180\deg$) or radial ejection
($\theta=0\deg$), the migration efficiency vanishes. In that case, the
product $a(m_\sta+m)$ of the semi-major axis with the total mass of
the system remains constant \citep{Hadjidemetriou_Icarus_1963}. Thus,
even if a Jupiter-mass planet orbiting a Sun-like star loses all its
mass, the semi-major axis increases only by a factor
$M_\Jup/M_\odot=0.1\%$.  The term $a\dot m/(m_\star+m)$ in
Eq.~(\ref{eq_dadt}) will thus be neglected in the
following.

For arbitrary migration efficiency, the solution of Eq.~(\ref{eq_dadt})
is ${a(t)} = a_0\left[1+\tau\ln\left( {m(t)}/{m_{0}}
\right)\right]^{-2}$. \bfx{For convenience, we provide an equivalent
relation, with orbital period instead of semi-major axis, deduced
from Kepler's third law
\be
\frac{P}{P_0} = \left(1+\tau\ln q\right)^{-3}\ ,
\label{eq_mp}
\ee
where $q=m/m_0$ is the fraction of remaining mass in the planet.} 
The expression (\ref{eq_mp}) is independent of time,
and holds for any mass-loss rate evolutions. The only requirement
concerns the efficiency of evaporation. \bfx{Here, we assume that the
evaporation process is sufficiently strong to decrease the mass of a
planet from $m_0$ to $m$.}

Using a typical ejection speed $V_\esc=15$ km.s$^{-1}$ and an
initial orbital velocity $v_0=145$ km.s$^{-1}$ corresponding to a
3 days orbit, the order of magnitude of the migration efficiency given
by (\ref{eq.tau}) is $\tau\sim V_\esc/v_0 = 0.1$. With this value, if a
Jupiter mass planet is transformed into a Neptune mass planet by
evaporation ($q=0.05$), the increase of its orbital period deduced from 
(\ref{eq_mp}) is $P/P_0 = 1.5$. For comparison, the ratio of the median
periods of Jupiter-like planets and Neptune-like planets (see
section~\ref{sec.observation}) is \bfx{4.9/3.4=1.4}.

\section{Observed period distributions}
\label{sec.observation}
From the more than 700 exoplanets discovered up to date, \bfx{204} have orbital
periods smaller than 10 days and mass smaller than 5
$M_\Jup$\footnote{Information extracted from Exoplanet Encyclopaedia,
http://exoplanet.eu/, on the  \bfx{29/11/2011}.}. One can group these planets in two
classes: Neptune-mass planets and smaller ($m < 2$ $M_\Nep$ or 2 $\times$
0.054 $M_\Jup$), and Jupiter-mass planets ($m>M_\Sat$ or 0.3 $M_\Jup$).
The first group contains \bfx{37} planets, from which we removed two couples
belonging to the same system\footnote{For the purpose of this study, we discard
all systems containing more than one planet with period shorter than 10
days. The removed planets are (CoRoT-7b, CoRoT-7c), (Gl581b, Gl581e),
(HD 40307b, HD 40307c), and \bfx{(Kepler-18b, Kepler-18c)}.}, while the
second \bfx{151}. Both correspond to 92\% of short-period planets considered.
The two populations are characterized by very distinct period
distributions: the Neptunes have a median period of \bfx{4.9} days, an average
period of \bfx{5.3} days and a standard deviation of \bfx{2.6} days, while for the
Jupiters the same quantities are of \bfx{3.4}, \bfx{3.5} and \bfx{1.6} days, respectively.
A Kolmogorov-Smirnov test shows that the cumulative distribution
functions (CDF) of the two distributions are significantly different,
with the probability that they were derived from the same parent
distribution being of \bfx{3.8}$\times10^{-5}$ (see Fig.~\ref{fig_histo}).

\prep{\figb}

It is important to keep in mind that the interpretation of these results
is conditioned by different factors, which warrant a discussion.
Firstly, the fitted parameters depend on the limits defined for the
planetary classes. However, the general properties of the distributions
are kept for a wide range of parameters, and the results are expected to
be resistant to changes of these parameters. More important is that the
observational bias imprinted on the sample by radial-velocity
detections, which depends on mass and period, is different for the
two classes of planets. Since the amplitude of a signal and thus its
detectability is proportional to $m P^{-1/3}$, heavier and
closer planets are detected more easily than lighter far away planets.
For this reason, the difference in median or average period values is
expected to increase for the real unbiased population, showing that the
difference in values we found here is not a detection effect but rather
an underestimation of the real one. In any case, the reflex motion
induced by a Neptune on a 10-day orbit on a 1 Solar-mass star is of 5
m.s$^{-1}$, significantly larger than the precision of state-of-the-art
spectrographs (e.g. HARPS) \citep{Mayor_Udry_PhST_2008}. As a
consequence, we expect the detections reported up to date to be
representative of the true population.

\section{Constraints and comparison with atmospheric models}
\label{sec.comparison}
\bfx{Here, we derive the distribution of the migration efficiency $\tau$
assuming that} the hot Neptune and the hot Jupiter populations
correspond to the same planets at different degrees of evaporation.
This provides constraints on the geometry of evaporation which are
then compared with results of atmospheric models. \bfx{That analysis
relies on the possibility that hot Jupiters can lose 95\% of their mass
by evaporation.} \bfx{Although additional works are required, we expect 
this} to happen at the early age of planet's
history, when stellar emission is strong \citep{Ribas_etal_ApJ_2005},
and planet density is low \citep[e.g.][]{Mordasini_etal_arxiv_2010}. Indeed,
these two conditions lead to important mass-loss
\citep{Erkaev_etal_AA_2007}.


In order to simplify the study, and to derive analytical expressions, we
approximate the CDFs of the two populations by log-normal distributions
\be
\PDF(\ln P_i) = {\cal N}_{\mu_i, \sigma_i}\left(\ln P_i\right)
= \frac{1}{\sqrt{2\pi \sigma_i^2}} \exp \left( 
\frac{(\ln P_i - \mu_i)^2}{2\sigma_i^2}\right)\ ,
\ee
where $i\in\left\{{\rm Jup}, {\rm Nep}\right\}$, and $P_i$ is expressed
in days. ${\cal N}_{\mu_i, \sigma_i}$ is the normal distribution.  The
best fits give $(\mu_\Jup = \bfx{1.19}, \sigma_\Jup = \bfx{0.39})$ for
Jupiter-like planets, and $(\mu_\Nep = \bfx{1.62}, \sigma_\Nep =
\bfx{0.57})$ for Neptune-like planets. Assuming that the migration
represented by $\Delta_{\ln P} = \ln P_\Nep-\ln P_\Jup$ is independent
of $P_\Jup$ and $P_\Nep$, the law of total probability implies that
$\Delta_{\ln P}$ follows a normal distribution with mean
$\mu_\Delta=\mu_\Nep-\mu_\Jup$ and standard deviation $\sigma_\Delta^2 =
\sigma_\Nep^2-\sigma_\Jup^2$.  Then, a change of variable from
$\Delta_{\ln P}$ to $\tau$ Eq.~(\ref{eq_mp}) gives
\be
\PDF(\tau) = -\frac{3 \ln q}{1+\tau \ln q}
{\cal N}_{\mu_\Delta, \sigma_\Delta}\left[-3 \ln\left(
1 + \tau \ln q\right)\right]\ .
\label{eq.pdf}
\ee
The PDF (\ref{eq.pdf}) is plotted in figure~\ref{fig_pdftau}(a) for
$q=0.03$, which is the ratio of the geometric mean of the
masses of the two planet populations. Using a typical ejection
speed $V_\esc=15$ km.s$^{-1}$ and the typical initial orbital velocity
$v_0=145$ km.s$^{-1}$, the PDF of $\tau$ expressed as a function of
$(\theta, \varphi)$ constrains the geometry of mass ejection (see
Fig.~\ref{fig_pdftau}(b)).

\prep{\figc}

In particular, if one considers an aperture $\varphi$ lower than
30$\deg$, one obtains the highest probable displacement in the eastward
direction at about 25$\deg$ of longitude (solid line in
Fig.~\ref{fig_pdftau}(b)). This value matches very well the one
predicted by atmospheric models \citep{Showman_Guillot_AA_2002,
Koskinen_etal_ApJ_2007, Showman_Polvani_arxiv_2011} or reported on HD
189733b \citep{Knutson_etal_nature_2007} (filled circle in
Fig.~\ref{fig_pdftau}(b)).

Nevertheless, it should be stressed again that the dynamical
models and the observation of the temperature map of HD 189733b do not
extend up to exobase. It is thus not possible to prove the connection
between the direction of mass ejection and the position of the hottest
region. 

\section{Conclusion}
\label{sec.conclusion}
\bfx{In this study, we provide a simple and very general analytical
expression of orbital migration induced by anisotropic evaporation.
The amplitude of this effect is not negligible.} Indeed, the order of
magnitude of the migration deduced from the expected ejection speed
$V_\esc$ and the typical orbital velocity $v_0$ of a 3-day orbit
is the same as the difference between the observed median period of hot
Jupiters and hot Neptunes. Moreover, the agreement is greatly enhanced
if ones assumes that the evaporation emanates from above the hottest 
region of the planet surface as measured on HD 189733b
\citep{Knutson_etal_nature_2007}, or predicted by atmospheric
circulation models \citep{Showman_Guillot_AA_2002, Koskinen_etal_ApJ_2007,
Showman_Polvani_arxiv_2011}. \bfx{Further studies on the dynamics
of the exobase may provide stronger constraints on the geometry of
evaporation. The analysis presented here would then be easily applied to
get precise evolution of evaporating planets.}

\begin{acknowledgements}
This work was supported by the European Research Council/European
Community under the FP7 through Starting Grant agreement number 239953.
We also acknowledge the support and from Funda\c{c}\~ao para a Ci\^encia
e a Tecnologia (FCT) through program Ci\^encia\,2007 funded by FCT/MCTES
(Portugal) and POPH/FSE (EC), and in the form of grant reference
PTDC/CTE-AST/098528/2008.

\end{acknowledgements}

\bibliographystyle{aa}
\bibliography{evaporation}

\end{document}